\documentclass[twocolumn,superscriptaddress,rmp,aps]{revtex4}
\usepackage{amsfonts,amssymb,amsmath,bm}
 \usepackage{graphicx}
\usepackage{url}
\usepackage{bm}

\newcommand{\1}{\mathrm{d}}

 \def\un{\hbox{{1\kern -0.25em\raise
0.4ex\hbox{{\scriptsize $|$}}}}} 
\def\nset{\hbox{{I\kern -0.18em N}}} 
   
\begin{document}


\title{Genesis of d'Alembert's paradox and analytical elaboration
of the drag problem} 

\author{G. Grimberg}
\email{gerard.emile@terra.com.br} \affiliation{Instituto  de
  Matem\'atica, Universidad  Federal do Rio de Janeiro (IM-UFRJ),
Brazil}
\author{W. Pauls}
\affiliation{Labor. Cassiop\'ee, UNSA, CNRS, OCA, BP 4229, 06304
Nice Cedex 4, France}
\affiliation{Fakult\"at f\"ur Physik, Universit\"at Bielefeld,
Universit\"atsstra{\ss}e 25, 33615 Bielefeld, Germany}
\author{U. Frisch}
\affiliation{Labor. Cassiop\'ee, UNSA, CNRS, OCA, BP 4229,
06304 Nice Cedex 4, France}
\draft{Physica D, in press; revised version} 

\begin{abstract}
We show that the issue  of the drag exerted by an incompressible fluid
on a body in 
uniform motion has played a major role in the early development of
fluid dynamics. In 1745 Euler came close, technically, to proving the 
vanishing of the drag for a body of arbitrary shape; for this he
exploited and significantly extended existing ideas on
decomposing the flow into thin fillets; he did not however have a 
correct picture of the global structure of the flow around a
body. Borda in 1766 showed that the principle of live forces implied
the vanishing of the drag and should thus be inapplicable to the
problem. After having at first refused the possibility of a
vanishing drag, d'Alembert  in 1768 established the paradox, 
but only for bodies with a head-tail symmetry. 
A full understanding of the paradox, as
due to the neglect of viscous forces, had to wait until the work of 
Saint-Venant in 1846. 
\end{abstract}



\maketitle



\section{Introduction}
\label{s:intro}

The first hint of d'Alembert's paradox -- the vanishing of the drag for a
solid body surrounded by a steadily moving ideal incompressible fluid -- had appeared even
before the analytical description of the flow of a 
``perfect liquid''\footnote{Kelvin's name of an incompressible inviscid
  fluid.} was solidly
established. Leonhard Euler in 1745, Jean le Rond d'Alembert in 1749 and 
Jean-Charles Borda
in 1766 came actually very close to formulating the paradox, using momentum
balance (in an implicit way) or energy conservation arguments, which actually predate its modern
proofs.\footnote{See, e.g. Serrin, 1959 and Landau and Lifshitz, 1987.}
D'Alembert in 1768 was the first to recognize the paradox as such, although in
a somewhat special case.  Similarly to Euler and Borda, his reasoning did not
employ the equations of motions directly, but nevertheless used a fully
constituted formulation of the laws of hydrodynamics, and exploited 
symmetries he had assumed for the
problem. A general formulation of d'Alembert's paradox for bodies of an
arbitrary shape was given in 1846 by Adh\'emar Barr\'e de Saint-Venant, 
who pointed out that the
vanishing of the drag can be due to  not taking into account viscosity. Other
explanations of the paradox involve unsteady solutions, presenting for
example a wake, as discussed by Birkhoff.\footnote{Euler,
1745; D'Alembert, [1749]; Borda, 1766; Saint-Venant, 1846, 1847; Birkhoff, 1950:~\S~9.}

Since the early  derivations of the paradox did not rely on Euler's
equation of ideal fluid flow, it was not immediatly recognized
that the idealized notion of an inviscid fluid motion was here
conflicting with the physical reality. The difficulties 
encountered in the theoretical treatment of the drag
problem were attributed to the lack of appropriate analytical
tools rather  than to any hypothetical flaws in the theory. 
In spite of the great achievements of Daniel and Johann Bernoulli, of d'Alembert and of Euler\footnote{See, e.g., Darrigol, 2005;
Darrigol and Frisch, 2007.} the theory of hydrodynamics seemed  beset with
insurmountable technical difficulties; to the contemporaries it thus appeared
of little help, as far as practical applications were concerned.
There was a dichotomy between, on the one hand,  experiments and the
everyday experience and, on the other hand  the eighteenth century's
limited understanding of the nature of fluids and of the theory of  
fluid motion. This dichotomy  is one of the
reasons why neither Euler nor Borda nor the early d'Alembert were able
to recognize and to accept the possibility of a paradox.

We shall also see, how the problem setting became more and more
elaborated in the course of time. Euler, in his early work on
the drag problem appeals to several physical examples of quite
different nature, such as that of ships navigating at sea and of
bullets flying through the air. D'Alembert's 1768 formulation of the
drag paradox is concrete, precise and much more mathematical (in the modern 
sense of the word) than Euler's early work. This is how
d'Alembert was able to show -- with much disregard for what
experiments or (sometimes irrelevant) physical intuition might
suggest -- that the framework of inviscid fluid motion necessarily
leads to a paradox.

For the convenience of the reader we begin, in Section~\ref{s:modern}, 
by recalling the modern proofs of
d'Alembert's paradox: one proof -- somewhat reminiscent of the
arguments in Euler's 1745 work -- relies on the calculation of the
momentum balance, the other one -- connected with Borda's 1766 paper --
uses conservation of energy. In Section~\ref{s:gunnery} we describe Euler's
first attempt, in 1745,  to calculate the drag acting on a body in a
steady flow using a modification of a method previously introduced
by D.~Bernoulli.\footnote{Bernoulli, 1736.}   
In Section ~\ref{s:resistance} we discuss d'Alembert's 1749
analysis of the resistance of fluids.  
In Section \ref{s:delucidationes} we review Euler's contributions to the drag
problem made after he had established  the equations of motions for ideal
fluid flow. 
Section~\ref{s:borda} is devoted to Borda's arguments against
the use of a live force (energy conservation) argument for this problem. 
In Sections
\ref{s:paradox} and \ref{s:venant} we discuss d'Alembert's and
St-Venant's formulation of the paradox. In Section \ref{s:conclusion}
we give the conclusions.

Finally, we mention here something which would hardly be necessary if 
we were publishing in a journal specialized in the history of science:
the material we are covering has already been dicussed several
times, in particular by such towering figures as Saint-Venant and
Truesdell.\footnote{Trusdell, 1954; Saint-Venant [1888].} Our contributions can only be considered
incremental, even if, occasionally, we disagree with our predecessors.

\section{Modern approaches to d'Alembert's paradox}  
\label{s:modern}

Let us consider a solid body $K$ in a steady potential flow with
uniform velocity $U$ at infinity. In the standard derivation 
of the vanishing of the drag\footnote{See,
e.g., Serrin, 1959.}  one proceeds as follows:
Let $\Omega $ be the domain bounded in the interior by the body $K$
and in the exterior by a sphere $S$ with radius $R$ (eventually, $R \to \infty
$). The force acting upon $K$ is calculated by writing a  momentum
balance, starting from the steady incompressible 3D Euler equation
\begin{equation}
{\bm v} \cdot \nabla {\bm v} = - \nabla p, \qquad  \nabla\cdot {\bm v}=0\;.
\end{equation}
The contribution of the pressure term gives the sum of the force
acting on the body $K$ and of the force exerted by the pressure on the
sphere $S$. It may be shown, using the potential
character of the  velocity field, that the latter force vanishes in  the limit
$R \to \infty $. The contribution of the advection
term can be written as the flux of momentum through the surface
of the domain $\Omega $: the flux through the boundary of $K$ vanishes
because of the boundary condition ${\bm v } \cdot {\bm n} = 0$; the
flux through the surface of $S$ vanishes because the velocity field is
asymptotically uniform ($ v \simeq U $ for $ R \to \infty $). From all this it
follows that the force on the body vanishes. This approach proves the vanishing
of both the drag and the lift.\footnote{The lift need not vanish if
  there
is circulation.}

Alternatively, one can use energy conservation to show the vanishing of the
 drag.\footnote{See, e.g., Landau and Lifshitz, 1987: \S~11.}  Roughly, the
 argument is that the work of the drag force, due to the motion with velocity
 $U$, should be balanced by either a dissipation of kinetic energy
 (impossible in ideal flow when it is sufficiently smooth) or by a flow to
 infinity of kinetic energy, which is also ruled out for potential flow.
This argument shows only the vanishing of the drag.

A more detailed presentation of such arguments may be found in the
book by Darrigol.\footnote{Darrigol, 2005:~Appendix~A.}

In the following we shall see that many technical aspects of these two
modern approaches were actually discovered around the middle of the eighteenth 
century.

\section{Euler and the New Principles of Gunnery (1745)}
\label{s:gunnery}

In 1745 Euler published a German translation of Robins's book ``New
Principles of Gunnery'' supplemented by a series of remarks whose
total amount actually makes up the double of the original volume. In
the third remark of the first proposition (Dritte
Anmerkung zum ersten Satz) of the 2nd Chapter  Euler
attempts to calculate the drag on a body at rest surrounded by a
steadily moving incompressible fluid.\footnote{For the German original of the third
  remark, cf. Euler, 1745: 259--270 (of \textit{Opera omnia} which we shall
  use for giving page references); an English
version, taken from  Hugh Brown's 1777 translation is available at
\url{www.oca.eu/etc7/EE250/texts/euler1745.pdf}\,. We shall sometimes
use our own translations.}

In 1745 the general equations governing ideal incompressible fluid
flow were still unknown. Nevertheless, Euler managed the remarkable
feat of  correctly calculating the force acting on an element of a
two-dimensional steady flow around a solid body.
For this, as we shall see, he borrowed and extended results obtained by 
D.~Bernoulli a few years earlier.\footnote{Bernoulli, 1736, 1738.}

\begin{figure}[h]

	\centering
		\includegraphics [width=8cm]{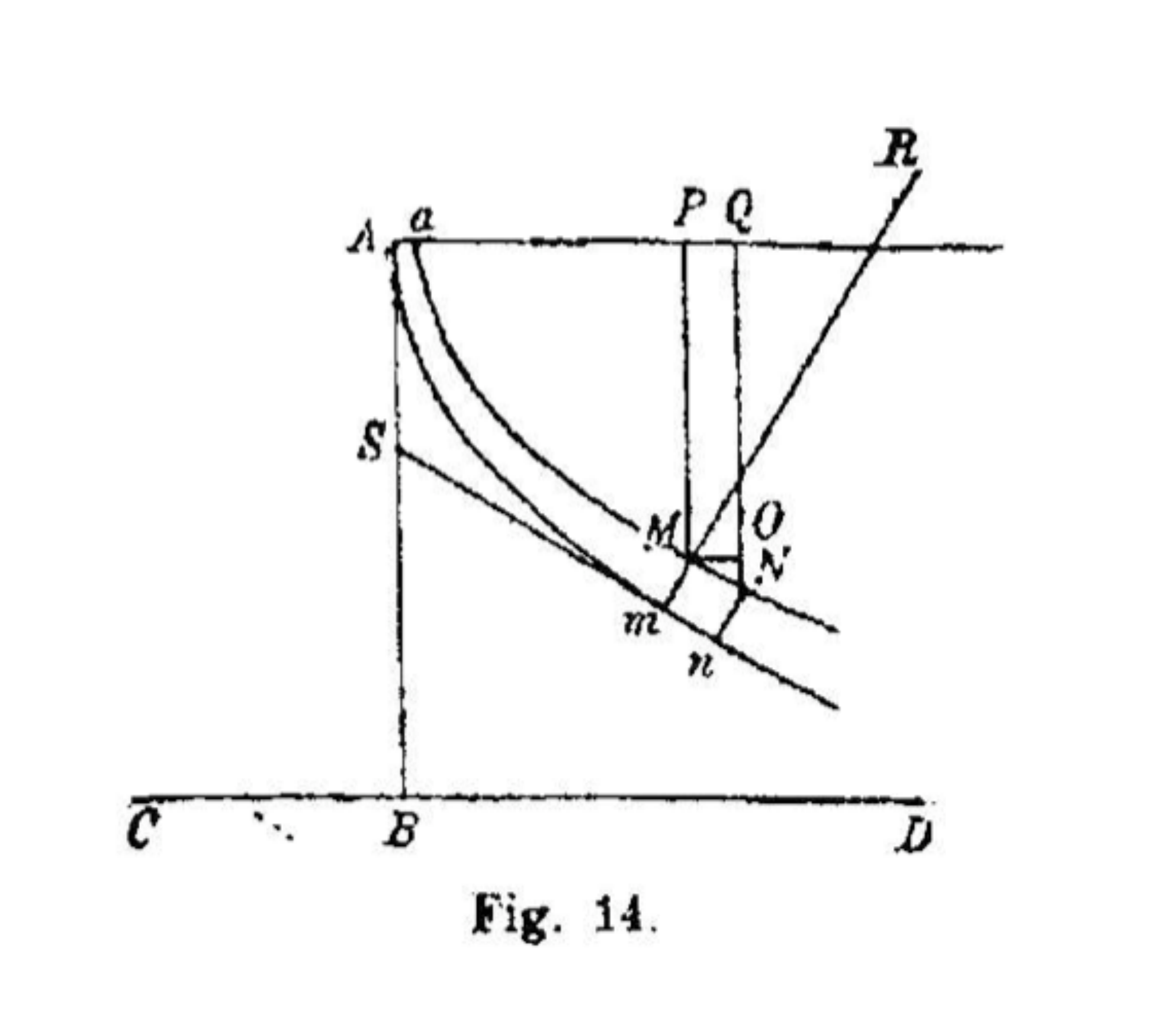}

	\caption{Figure 14 of Euler, 1745:~263: this figure represents
	a  fillet of fluid aAMm, deflected by the solid body, 
but the shape of the body is not fully  specified.
	}
	\label{fig:file1}
\end{figure}

Euler begins by noting that instead of calculating the drag acting on
a body moving in a fluid one can calculate the drag acting on a
resting body immersed in a moving fluid. Thus, he considers a fluid
moving into the direction AB\footnote{Here, contrary to the usage in
  Eulers' memoir, all geometrical points
  will be denoted by roman letters, leaving italics for algebraic
  quantities.}  (cf. Figs.~\ref{fig:file1} and \ref{fig:file3}),
past a solid body CD.\footnote{These are Euler's
own words; examination of various of his figures and of the scientific 
context shows that the body extends below CD and, perhaps also above.}
Then Euler continues by describing the motion of fluid particles and
establishes a relation between the trajectory and velocity of each
fluid particle and the
force which is acting on this particle. He observes that, instead
of determining the force on the body, one can evaluate the reaction on 
the fluid:

{\footnotesize
\noindent 
	But since all parts of the fluid, as they approach the body,
	are deflected and change both their speed and direction [of
	motion], the body has to experience a force of strength equal
	to that needed for this change in speed as well as direction
	of the particles.\footnote{Euler, 1745:~263. Weil aber alle Theile der
	fl\"u{\ss}igen Materie, so bald sich dieselben dem K\"orper
	nahen, gen\"othiget werden auszuweichen, und so wohl ihre
	Geschwindigkeit, als ihre Richtung zu ver\"andern, so mu{\ss}
	der K\"orper eine eben so gro{\ss}e Kraft empfinden, als zu
	dieser Ver\"anderung so wohl in der Geschwindigkeit, als der
	Richtung der Theilchen, erfordert wird.}}\\[-1.ex]

Thus, one has to determine the force which is applied at each point of
the fluid. 
Euler chooses a fillet\footnote{Euler uses the word
  ``Canal'' (channel).} AaMm of fluid with an infinitesimal
width  and observes that the velocity\footnote{Following early eighteenth
 century notation, Euler represents a velocity by the corresponding height
of free fall to achieve the given velocity, starting a rest; in modern
notation this would be $\sqrt{2gh}$. In the 1745 paper
Euler takes mostly $g=1$ -- but  occasionally
$g=1/2$ -- and denotes the height by $v$. In order not to confuse the
reader, we shall here partially
modernize
the notation and in particular  denote the velocity by $v$.} $v$ of the particles
passing 
through the section Mm is
inversely proportional to its  (infinitesimal) width $Mm = \delta
z$; so that $v\, \delta z=v_0\, \delta z_0 $, 
where $\delta z_0={\rm Aa}$ and $v_0$ are the width of the fillet and the
velocity  
at the reference point A.\footnote{Euler denotes our $\delta z$,
$\delta z_0$ and
$v_0$ by $z$,
$a$ and  $\sqrt{2b}$, respectively.} For later reference let us call this relation mass
conservation.
Euler  assumes that the particles
passing through the section Aa follow the fillet AaMm. This is
equivalent to assuming that the velocity in each section Mm along
the trajectory depends only on the location of the point M and not on
time,
in modern terms a stationary flow. Here appear for the first time
explicitly the concepts of  streamline and of stationarity in two dimensions.

With the above assumption, Euler defines
\begin{equation}
\begin{split}
& AP=x, \quad PQ=\1x, \quad PM=y, \quad ON=\1y, \\
& p=\1y/\1x, \quad
MN=\sqrt{\1x^2+\1y^2}=\1x\sqrt{1+p^2}\;. 
\end{split}
\label{vardef}
\end{equation}
Since the force exerted by the body on the fluid is
oriented upward, we prefer  orienting the vertical
axis upward. Hence $y$ and $p$ will be negative in what follows. 
Otherwise we shall
mostly follow Euler's notation.
Euler calculates the normal 
and tangential components, $\1F_{\rm
  N}$ and $\1F_{\rm T}$, of the infinitesimal force acting on the element of 
fluid 
fillet  MNnm (see Fig.~\ref{fig:file1}).\footnote{The notation $\1F_{\rm N}$ and $\1F_{\rm T}$ is ours.}

With the assumed unit density, the mass of fluid in  MNnm is
\begin{equation}
\delta z \times MN=\delta z\1x\sqrt{1+p^2}\; .
\label{themass}
\end{equation}

The normal  acceleration $\1F_{\rm N}$ in the direction
MR
is calculated by Euler as a centripetal
acceleration, i.e., given by the product of the square of the
velocity $v^2$ and the curvature $\left(1+p^2\right)^{\frac{3}{2}}\1p/\1x$. 
Euler may  here by following  D.~Bernoulli.\footnote{Bernoulli
Daniel, 1736 and 1738: Section~XIII, \$~13.} The latter, in a paper 
concerned among other things  with jets impacting on a plane, had developed an 
analogy between an
element
of fluid following a curved streamline and a point mass on a curved
trajectory (cf. Fig.~2).
\begin{figure}[h]

	\centering
		\includegraphics [width=5cm]{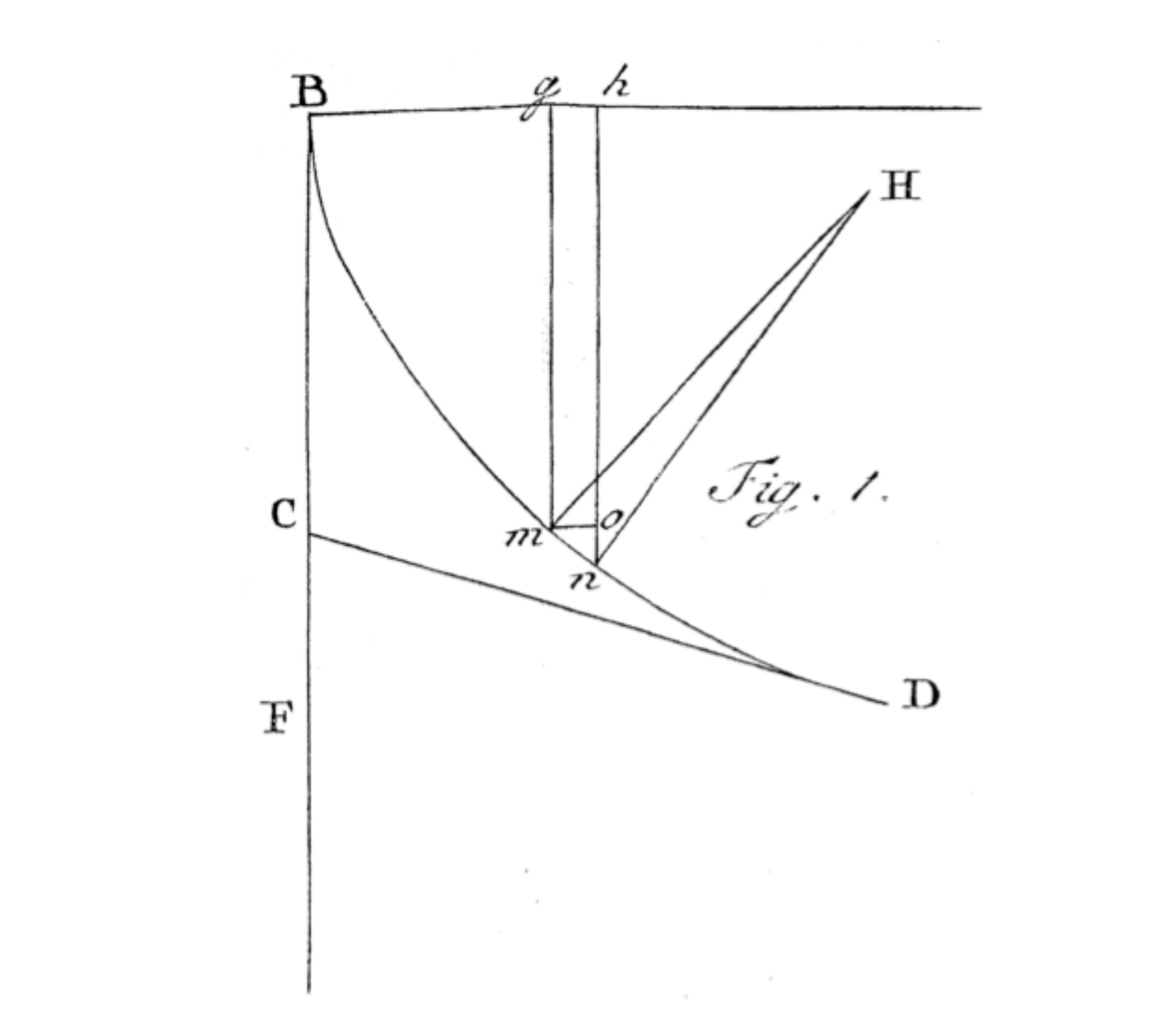}
\label{fig:file2}
\caption{
Figure 1  of Bernoulli, 1736. A centripetal argument is used to
calculate
the normal force acting along a  fillet of fluid represented here
just by the curve BD (changes in width are ignored).}
\end{figure}
Multiplying the acceleration by the elementary mass and using mass 
conservation,\footnote{Bernoulli, 1738:~287 assumed a fillet of uniform
  width (\textit{fistulam implantatam esse uniformis quidem amplitudinis}) 
and thus did not use mass conservation to
  relate
the varying width and  velocity.} 
 Euler then obtains
\begin{equation}
\1F_{\rm N}=v_0\delta z_0v\1p/(1+p^2)\;,
\label{exprfn}
\end{equation}
in which the velocity $v$ along the fillet is considered to be a
function of the slope $p$.
 
To obtain the tangential force $\1F_{\rm T}$ 
in the direction  mS on the element of
fillet,  Euler  writes 
\begin{equation}
\delta  z\1x\sqrt{1+p^2}\1( v^2/2) = -\1F_{\rm T}\ \1x\sqrt{1+p^2}\;,
\label{isitliveforce}
\end{equation}
and thus
\begin{equation}
\1F_{\rm T} = - \delta z \1(v^2/2) = -\delta z_0 (v_0/v) \1(v^2/2)\;.
\label{exprft}
\end{equation}
For the case of Fig.~\ref{fig:file1} the force is oriented in the
direction mS, because the fluid is moving more slowly at N than at M.
Euler does not elaborate on how he derives \eqref{isitliveforce} but
this
seems typically a ``live force'' argument of a kind frequently used
at that time, for example by the Bernoullis.\footnote{Cf., e.g., Darrigol, 2005: Chap.~1.} Indeed the l.h.s is the
variation of the live force (kinetic energy) and the r.h.s. 
is what we would now call the work of the tangential force per unit mass.

So as to later  determine the drag, that is the force on the body in
the vertical direction,  Euler adds these normal and
tangential elementary forces,  projected onto the vertical axis oriented in the
direction BA. He thus obtains the following elementary vertical force
on the fluid:
\begin{equation}
\1F_{\rm BA}=v_0\delta z_0\left( \frac{v\1p}{(1+p^2)^{\frac{3}{2}} } + 
\frac{p\1v}{\sqrt{1+p^2} } \right)\; .
\label{elemvert}
\end{equation}
Here a ``miracle'' happens:  the r.h.s. of \eqref{elemvert} is the exact
differential of
\begin{equation}
v_0\delta z_0\left(\frac{vp}{\sqrt{1+p^2}}\right)\;.
\label{miracle}
 \end{equation}
Finding the exact form of the function $v(p)$,  as we now know,
requires
the solution of  a non-trivial boundary value problem. The exact form
does however not matter for the integrability property and -- from
a modern perspective -- can 
be related to the global momentum conservation property of the
Euler equation. In 1745 Euler did not comment on the miracle.
It is worth stressing that it does not survive if any error is made regarding
the numerical factors appearing in the normal
and tangential force.

Euler is now able to exactly integrate the elementary force along a fillet
from its starting point A, assumed to be far upstream ($p = -\infty$), to a 
point m with a finite slope $p$. Noting that $-p/\sqrt{1+p^2}$ is the
cosine
of the angle  $MSB$, he obtains the following force on the body, due 
to the fillet:
\begin{equation}
F_{\rm AB}=-v_0^2 \delta z_0\left(1-\dfrac{v}{v_0}\cos MSB\right)\;.
\label{forceonfillet}
\end{equation} 

Note that this is a force from a given fillet of infinitesimal width
which must still be integrated over a set of fillets encompassing the
whole fluid. More important here is where to terminate the fillet. It
is clear that the relevant fillets start far upstream in the vertical
direction;
but where do they lead after having come close to the solid body?
Euler considers various possibilities, such as a $90^\circ$
deflection.
He then envisages a very interesting case:

{\footnotesize
\noindent It remains therefore only to fix upon the point
which is to be esteemed the last of the canal. If we go so far that
the fluid may pass by the body, and attain its first direction and
velocity then shall $\delta z= \delta z_0$, and the angle mSB vanish,
and therefore 
its
cosine $=1$, then shall the force acting on the body in the direction
AB $=-v_0^2\delta z_0 (1-1) =0$, and the body suffers no resistance.\footnote{
Euler, 1745:~267. Hier k\"ommt es also nur darauf an, wo das Ende des
Canals angenommen werden soll. Geht man so weit, bi{\ss} die fl\"u{\ss}ige Materie um den K\"orper
v\"ollig vorbey geflossen, und ihren vorigen Lauf wiederum erlanget
hat, so wird \dots, und der Winkel mSB verschwindet, dahero der
Cosinus desselben $=1$ wird. In diesem Fall w\"urde also die auf den
K\"orper
nach der Direction AB w\"urkende Kraft \ldots und der K\"orper litte
gar keinen Wiederstand.}}\\[-1.ex]

From a technical point of view Euler's 1745 derivation of the vanishing
of the drag force has many features of the modern proof. However Euler
refuses here to see a paradoxical property of the model of ideal
fluid flow (for which the equations are not even completely
formulated). He accepts the possibility that the vanishing of the drag
applies to certain exotic fluids wich are ``infinitely fluid \ldots  and also
compressed by an infinite force''\footnote{Euler, 1745:~268--269. 
\ldots unendlich fl\"u{\ss}ig 
\ldots 
und von einer unendlichen Kraft zusammen gedruckt \ldots} 
such as the hypothetical ether surrounding celestial bodies 
(called by him ``subtle heavenly material''), but he firmly rejects it
for water and air. Indeed, immediately after the previous citation he
writes:

{\footnotesize
\noindent  Hence it appears, that for air or water, we are not to take
the point of the canal for last, where the motion behind the body
corresponds exactly with that at the beginning of the canal.\footnote
{Euler, 1745:~267. Woraus erhellet, da{\ss} man f\"ur Wasser und Luft nicht denjenigen
Punkt des
Canals, wo die Bewegung hinter dem K\"orper mit der ersten wiederum
v\"ollig \"ubereinkommt, f\"ur den letzten annehmen
k\"onne.}}\\[-1.ex]

Euler then explains why in his opinion the ``last point'' should not
at all
be taken far downstream, but rather near the inflection point M  where
the angle MSB achieves its maximum value, as
shown in Fig.~\ref{fig:file3}.\footnote{Truesdell, 1954:~XL writes that
``Euler supposes that the oncoming fluid turns away from the axis, leaving
a dead-water region ahead of the body''; actually,  Euler does not
assume  any
dead-water region in his Third Remark.}
\begin{figure}

	\centering
		\includegraphics [width=5cm]{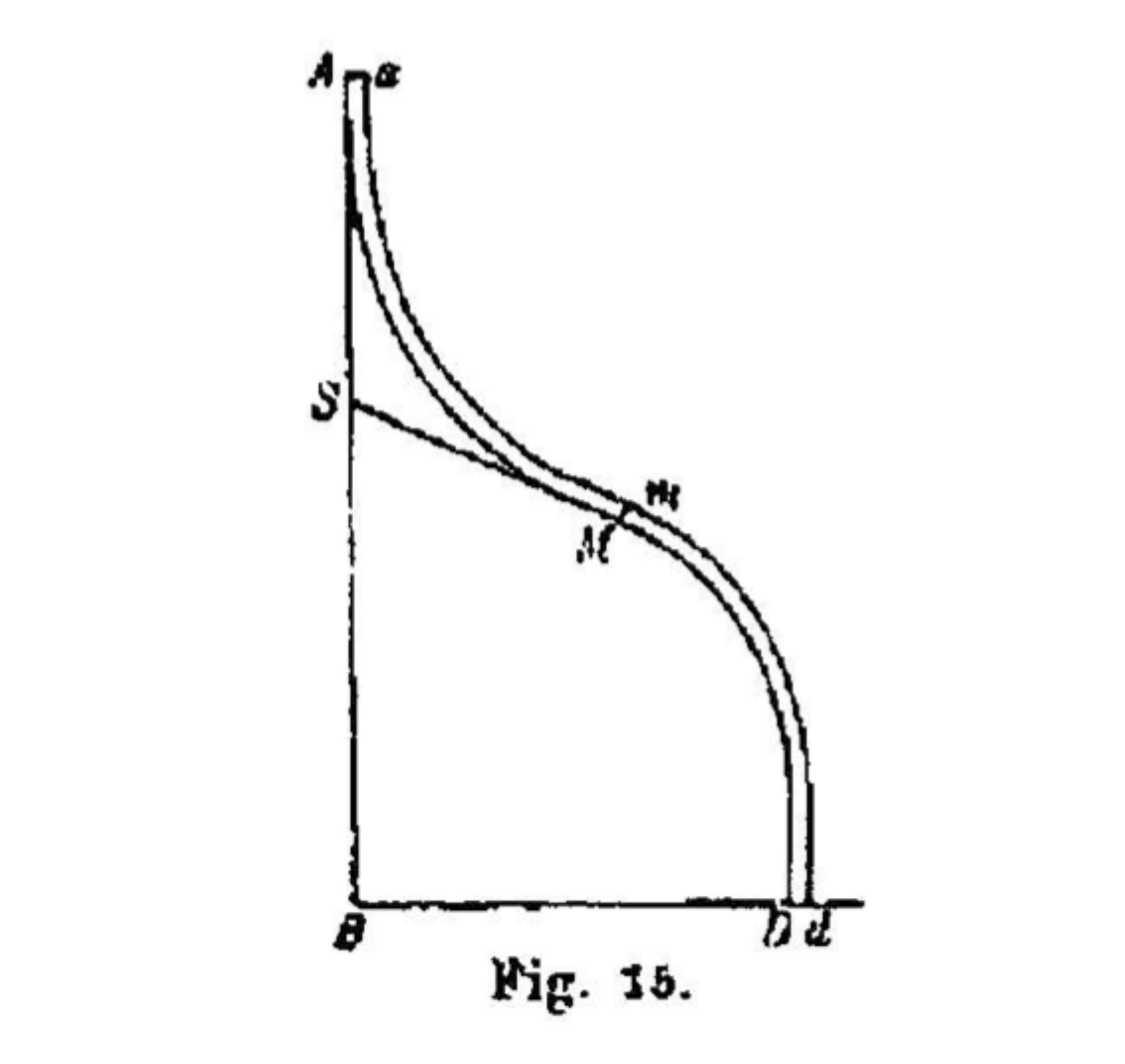}

	\caption{Figure 15 of Euler 1745:~268 from which he tries to
	explain that the drag should be calculated using only the
	portion AM of the fillet.}
	\label{fig:file3}
\end{figure}
As pointed out to us by Olivier Darrigol, in Euler's opinion the portion AM of
the canal AD is the only one that exerts a force on the body, the alleged
reason being that the force caused by the deflection in the portion MD is not
directed toward the body:

{\footnotesize
\noindent The other part DM produces a force which is opposite to the first,
and would cause the body to move back in the direction BA. Now, as only a true
pressure [a positive one] can set a body into motion, the latter force can
only act on the body insofar as the pressure of the fluid matter from behind
is strong enough to move the body forwards.\footnote {Euler, 1745:~268. Aus
dem andern Theil DM aber ensteht eine Kraft, welche jener entgegen ist, und
von welcher des K\"orper nach der Direction BA zur\"uck gezogen werden
sollte. Da nun kein K\"orper anders, als durch einen w\"urklichen Druck in
Bewegung gesetzt werden kann, so kann auch die letztere Kraft nur in so ferne
auf den K\"orper w\"urken, als der Druck der fl\"u{\ss}igen Materie von
hinten stark genug ist, den K\"orper vorw\"arts zu stossen.}}\\[0.6ex] Hence he
departed from strict dynamical reasoning to follow a dubious intuition of the
transfer of force through the fluid.\footnote{Darrigol,
private communication, 2007.}

To sum up, Euler performed a real \textit{tour de force} by deriving the
correct expression for the force on a fillet of fluid without having the
equations of motion but practically he was not able to reach much beyond
Newton's impact theory when considering the global interaction between the
fluid and the body.

\section{D'Alembert and the Treatise on the resistance of fluids (1749)}
\label{s:resistance}

In a treatise\footnote{D'Alembert, [1749], 1752.} written
for the prize of the Berlin Academy of 1749 whose subject was the
determination of the drag a flow exerts upon a body, d'Alembert
gives a description of the motion of the fluid analogous to that of
Euler.  It is not clear if d'Alembert knew about Euler's ``Commentary 
on Gunnery''. As noted
by Truesdell\footnote{Truesdell, 1954:~LII.}, some
figures in d'Alembert's treatise are rather similar to those found in
the  Gunnery but there are also arguments in the Gunnery which would
have allowed d'Alembert, had he been aware of them, 
to extend his 1768 paradox to cases not
possessing the head-tail symmetry he had to assume. Anyway, d'Alembert was
fully aware of D.~Bernoulli's work on jet impact in which, as we already
pointed out,  a similar figure is found. 

In the treatise d'Alembert described  the motion of an incompressible fluid in uniform
motion at large distance, interacting with a localized axisymmetric
body. He observed  that the streamlines and the
velocity of the fluid at each point in space are time-independent. The
velocity $a$ of the fluid far upstream of the body is directed along the
axis of symmetry (which he takes for the abscissa); the other axis is
chosen to be perpendicular to this direction. In this frame a point
M of the fluid is characterized by the cylindrical coordinates $(x,z)$ and the 
corresponding velocity has the components $av_x$ and 
$av_z$.\footnote{D'Alembert uses a similarity argument to prove that
the velocity field around a body of a given shape is proportional to
the incoming velocity $a$ (d'Alembert, [1749]:~\S42--43, 1752:~\S 39).}

D'Alembert's first aim is  to derive the partial differential
equations which determine the motion of the fluid, and the appropriate
boundary conditions with which they must be supplemented.  He observed
that, in order to  determine the drag on the body, one must first
determine

{\footnotesize
\noindent 
        \ldots the pressure of the fillet of Fluid which glides
         immediately on the surface of the body. For this it is
        necessary to know the velocity of the particles of the
        fillet.\footnote{D'Alembert, 1752:~xxxi. \ldots la pression du filet de Fluide qui
        glisse imm\'ediatement sur la surface du corps. Pour cela il
        est n\'ecessaire de conno\^itre la vitesse des particules de
        ce filet.}}\\[-1.ex]
%
 
By considering the motion of fluid particles during an infinitesimal
time
interval, d'Alembert is able to find the expressions of the two
components of the force acting on an element of fluid:
\begin{equation}
\gamma_z
  =a^2\left(-v_x\frac{\partial v_z}{\partial x}-v_z\frac{\partial
  v_z}{\partial z}\right)\; , 
\end{equation}
and 
\begin{equation}
 \gamma_x= a^2\left(-v_x\frac{\partial
  v_x}{\partial x}-v_z\frac{\partial v_x}{\partial z}\right)\;.
\end{equation}
From this d'Alembert derived for the first time the partial differential equations
for axisymmetric, steady, incompressible and irrotational
flow, but he does not use such equations in considering the problem of
``fluid resistance''.\footnote{Cf. Truesdell, 1954:~LIII, Grimberg, 1998:~44--46, 
Darrigol, 2005:~20--21.}

How does d'Alembert
calculate the drag? From an assumption about the continuity
of the velocity he infers, contrary to Euler, that there must be a zone of stagnating
fluid in front of the body and behind it, bordered by the streamline
TFMDLa which attaches to the body at M and detaches at L (see
Fig.~4).\footnote{D'Alembert,   [1749]:~\S~39,
1752:~\S~36.}

In his calculation of the drag d'Alembert used an approach which
differed from that of Euler in the Gunnery: instead of calculating the 
balance of
forces acting on the fluid he considered the pressure force exerted 
on the body by the fluid fillet in immediate contact with it. 
D'Alembert noted first that, for each surface element of the body,  the force exerted by the
fluid  particles is perpendicular to this surface, because
of the vanishing of the tangential forces. characterizing the flow of an
ideal  fluid.\footnote{D'Alembert, [1749]:~\S~40; 1752:~\S~37. This vanishing,
as we know, characterizes an ideal fluid;  d'Alembert did not relate
it to the nature of the fluid.}

In conformity with Bernoulli's law, d'Alembert  expressed the pressure along the
body as $a^2(1-v_x^2-v_z^2)$. With $\1s$
denoting the element of curvilinear length along the sections of the
body
by an axial plane such as that of Fig.~4), the infinitesimal
element of surface of revolution of the body upon which this pressure
is acting is $2\pi z\1s$. 
\begin{figure}[h]

	\centering
\includegraphics [width=5cm]{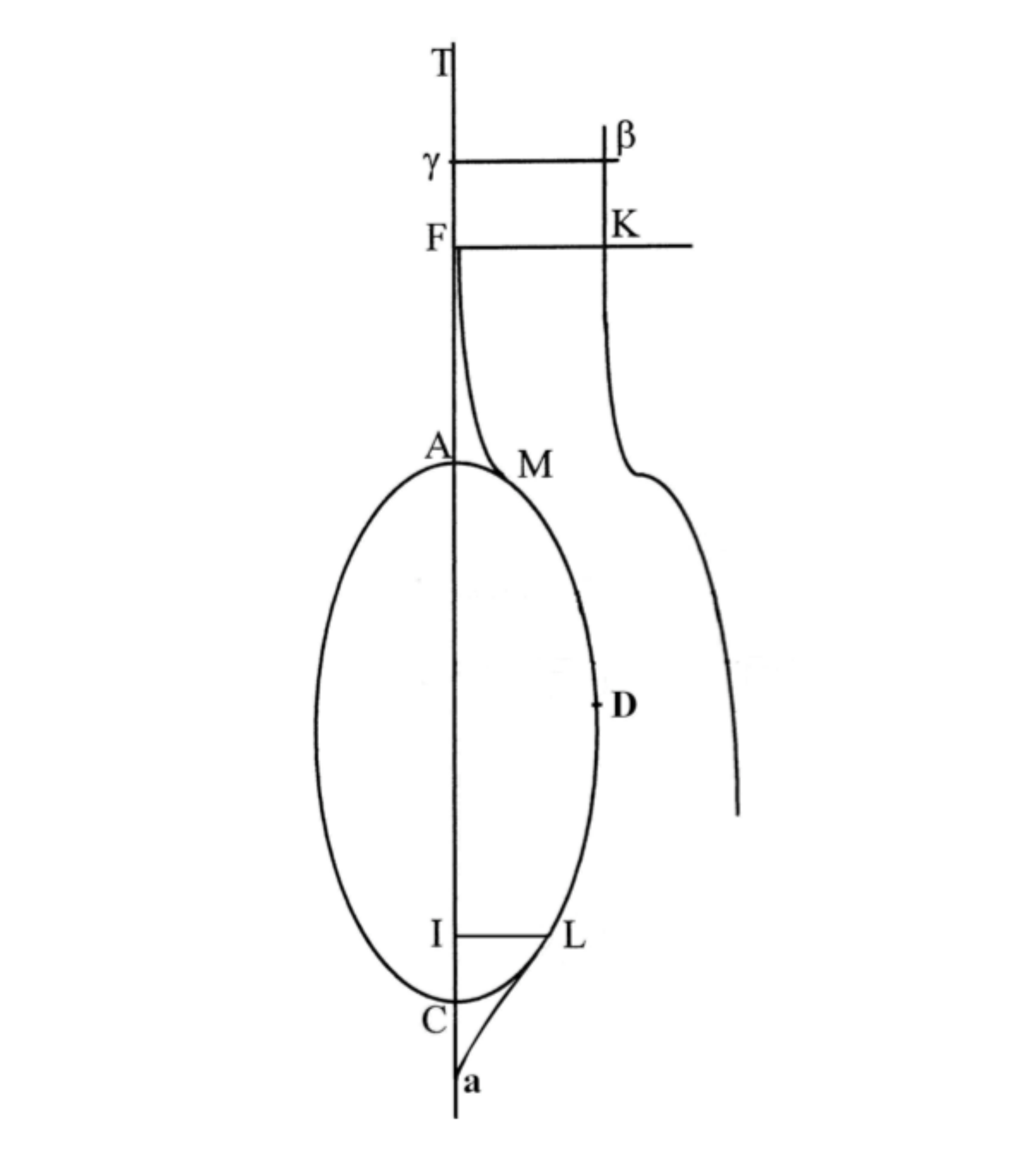}
	\label{fig:file4}
	\caption{Figure 14 of d'Alembert, [1749] redrawn. Not all elements
shown here are  used in our arguments.}
\end{figure}
The component along the axis of the pressure
force exerted  is 
\begin{equation}
2\pi a^2(1-v_x^2-v_z^2)z \1z\; .
\end{equation}
Further integration along the profile AMDLC yields
the vertical component of the drag.

Then came a very important remark.  D'Alembert noted that in the case
of a body which is not only axisymmetric but has a head-tail
symmetry,\footnote{In d'Alembert, 1752 this additional symmetry is
explicitly assumed; in d'Alembert, [1749] the language used only
suggests such a symmetry.}  the contributions to the drag from two
symmetrically located points would be equal and of opposite sign and
thus cancel.\footnote {D' Alembert, [1749]:~\S~62,
1752:~\S~70.} In order to avoid the vanishing of the drag, he assumed
that the attachment point M and  the separation point L are not
symmetrically located:

{\footnotesize
\noindent
 From there it follows that the arcs LD and DM cannot be
	equal; because, if they were, the quantity $-\int{2\pi
	y\1y(p^2+q^2)}$ would be equal to zero so that the body would
	not experience any force from the fluid: which is contrary to
	experiments.\footnote{D'Alembert, 1752:~\S~70. Del\`a il s'ensuit que les arcs LD, DM
	ne sauroient \^etre \'egaux\,; car s'ils l'\'etoient, alors la
	quantit\'e $-\int{2\pi y\1y(p^2+q^2)}$ seroit \'egale \`a
	z\'ero de mani\`ere que le corps ne souffriroit aucune
	pression de la part du fluide\,: ce qui est contre
	l'exp\'erience.}}\\[-1.ex]

This stress on ``experiments'', already present in the 1749 manuscript
and which will not reappear in d'Alembert's 1768 paradox paper, seems
to reflect just common sense. It cannot be explained by d'Alembert's
hypothetical desire to adhere to late recommendations by the Berlin
Academy which emphasized comparisons with experiments for the 1750
prize on resistance of fluids. D'Alembert did not seem pleased with
such late changes and these recommendations were probably formulated
only in May 1750.\footnote{D'Alembert, 1752:~xxxviii; Taton and
Yushkevich, 1980:~312--314; Grimberg, 1998:~9.}

D'Alembert's new idea, compared to Euler, is to consider the
drag as the resultant of the pressure forces directed along the normal 
to the surface of the body over its entirety. But for d'Alembert it is still
unimaginable to obtain a vanishing drag. 
%

\section{Euler  and the `Dilucidationes' (1756)}    
\label{s:delucidationes}
%

The \textit{Dilucidationes de resistentia fluidorum} (Enlightenment
  regarding the resistance of fluids) have been written in 1756, one year 
after Euler established his famous equations in their final 
form.\footnote{Euler, 1755, 1756.} In his review of previous efforts
to understand the drag problem for incompressible fluids, Saint-Venant\footnote{Saint-Venant,
  [1888], probably mostly written around 1846.} writes the following
  about the \textit{Dilucidationes}:

{\footnotesize
\noindent
And it is obvious that, when the flow is assumed indefinite or very
broad, the theory of the \textit{Dilucidationes} can only be and
actually is
just a return to the vulgar theory, \ldots\footnote{Saint-Venant, [1888]:~35.
Et il est 
\'evident que, lorsque le courant est suppos\'e
ind\'efini ou tr\`es large, la th\'eorie des \textit{Dilucidationes} 
d'Euler ne peut \^etre et n'est r\'eellement qu'un retour pur et
simple \`a la th\'eorie vulgaire,}}\\[-1.ex]

Here, Saint-Venant understands by ``vulgar theory'' the  impact
theory which goes back to the seventeenth century. Actually, in 1756 Euler
was rather pessimistic regarding the applicability of his equations
to the drag problem:

{\footnotesize
\noindent
But the results which I have presented in several previous
	memoirs on the motion of fluids do not help much
	here. Because, even though I have succeeded in reducing
	everything that concerns the motion of fluids to analytical
	equations, the analysis has not reached the sufficient degree
	of completion which is necessary for the solving of such
	equations.\footnote{Euler, 1760:~200. Quae ego etiam nuper in aliquot
	dissertationibus de motu fluidorum exposui, nullum subsidium
	huc afferunt. Etiamsi enim omniam quae ad motum fluidorum
	pertinent, ad aequationes analyticas reduxi, tamen ipsa
	Analysis minime adhuc ita est exculta, ut illis aequationibus
	resoluendis sufficiat. Euler, 1760, \S~6:~200.}}\\[-1.ex]

Truesdell discusses the \textit{Dilucidationes} in detail.\footnote{
  Truesdell 1954: C--CVII.}  Actually this paper is quite famous
because of a remark Euler made on the cavitation that arises from 
negative pressure in incompressible fluids. Truesdell is also rightly
impressed by Euler's  success in doing something non-trivial with his equation
for flows around a  parabolic cylinder; for this Euler uses a system
  of
 curvilinear coordinates based on the streamlines and their orthogonal
  trajectories.

 The \textit{Dilucidationes} are however not contributing much to our 
understanding of drag. In \S~15, Euler expresses his doubts regarding
the applicability of his 1745 calculation to both the front and
the back of a body (which would result in vanishing drag):

{\footnotesize
\noindent \ldots the
boat would be slowed down at the prow as much as it would be pushed at
the poop \ldots\footnote{Euler, 1760:~206....puppis nauis paecise tanta vi propelleretur,
quanta prora repellitur...}}\\[-1.ex]

We must mention here that, because of a possible non-vanishing transfer of
kinetic energy to infinity, the modern theory of the d'Alembert paradox does
not apply to flow with a free surface, such as a boat on the sea.

Thus, in the \textit{Dilucidationes}  we find a first attempt to
introduce 
a new
analytical treatment of streamlines unrelated to the previous theories and
coming closer to the modern description of a fluid
flow. Nevertheless, Euler does not succeed in using his 1755 equations
to improve our understanding of the drag problem.

\section{Borda's memoir (1766)}  
\label{s:borda}

In his memoir Borda, a prominent French ``Geometer''
and experimentalist,  studies the loss of ``live force''
(energy) in incompressible flows,
in particular in pipes whose section is abruptly
enlarged.\footnote{Borda, 1766.} At the end of his memoir Borda
gives an example of what would be, in his opinion, ``a bad use'' of the
principle of conservation of live forces. This is precisely the 
problem of determining the
drag force that a moving fluid exerts upon a body at rest. The
particles of the fluid in the neighborhood of the body ``delineate
curved lines or rather move in small curved channels''; the pressure
force acting upon the body has to be determined. But the channels
become narrower at certain locations similarly to a siphon, so that
the principle of live forces  cannot be used. To prove this point
he then presents  the following argument for the vanishing of the drag:

%

{\footnotesize
\noindent
\ldots suppose that the body $D$ moves uniformly through a quiescent fluid,
driven by the action of the weight $P$. According to this principle [of live
forces], the difference of the live force of the fluid must be equal
to the difference of the actual descent of the weight; however, since
the motion is supposed to have reached uniformity, the difference of
the live forces equals zero. Therefore, the difference of the actual
descent is also zero, which cannot happen unless the weight $P$ is
itself zero. As the weight $P$ measures the resistance of the fluid,
the supposition of the principle [of live forces] necessarily leads to a
vanishing resistance.\footnote{Borda,
1766:~604--605. \ldots supposons que le corps $D$ se meuve
uniform\'ement dans un fluide tranquille, entra\^{\i}n\'e par l'action
du poids $P$\,: on sait que suivant le principe, la diff\'erence de la
force vive du fluide devra \^etre \'egale \`a la descente actuelle du
poids $P$\,; mais puisque le mouvement est cens\'e parvenu \`a
l'uniformit\'e, la diff\'erence des forces vives $=0$\,; donc la
diff\'erence de la descente actuelle sera aussi $=0$, ce qui ne se
peut pas \`a moins que le poids $P$ ne soit lui-m\^eme $=0$\,: or le
poids $P$ marque la r\'esistance du fluide\,: donc la supposition du
principe dont il s'agit, donne toujours une r\'esistance nulle.}}\\[-1.ex]


This constitutes the first derivation of the d'Alembert paradox using
an energy dissipation argument. Borda's explanation of why the live
force conservation argument is inapplicable rests on the
aforementionned analogy with the siphon problem. This is illustrated
by a figure\footnote{Borda, 1766:~Fig.~14, found at the end of the 1766
  volume on p.~847.} not reproduced here because of its poor quality.
There one sees a fillet of fluid narrowing somewhat as it approaches
the body. The modern concept of dissipation in high-Reynolds-number
flow
being confined to regions with very strong velocity gradients is
definitely
not what Borda had in mind.


%

Borda's reasoning is  correct, but like Euler in 1745 and d'Alembert
in 1749, he does not formulate the vanishing of the drag as a paradox.
In his remarks Borda addresses neither the question of the nature
of the fluid, nor the consequences of having stationary streamlines,
nor the problem of the contact between the fluid and the body (absence
of viscosity in the case of ideal flow) which, as we know, are quite
central to the understanding of the paradox.
%

\section{D'Alembert's memoirs on the paradox (1768 and 1780)}  
\label{s:paradox}

In Volume V of his ``Opuscules'' published in 1768, a part of a
memoir  is entitled ``Paradox on the resistance of fluids proposed
to geometers.''\footnote{D'Alembert, 1768. In the eighteenth century
  ``Geometer'' was frequently used to mean ``mathematician'' (pure or applied).} 
%

D'Alembert considers again an axisymmetric body, but now  with a
head-tail symmetry. More precisely, he assumes a plane of
symmetry perpendicular to the direction of the incompressible flow at large distance
and dividing the body into two mirror-symmetric pieces.
To avoid the problem of possible separation of streamlines
upstream and downstream of the body, he assumes that the front part
and the rear part of the body have needle-like endings. First of all he asserts
that the velocities at every location in the fluid are perfectly
symmetric in front/rear of the body, and that

{\footnotesize
\noindent
        \ldots under this assumption the law of the equilibrium and the
        incompressibility of the fluid will be perfectly obeyed,
        because, the rear part of the body   being similar and equal to its
        front part, it is easy to see that the same values of $p$
        and $q$ [i.e.\ the velocity components] which will give at the
        first instant the equilibrium and incompressibility of the
        fluid at the front part will give the same results for the
        rear part.\footnote{D'Alembert, 1768:~133. \dots dans cette supposition les loix de
        l'\'equilibre \& de l'incompressibilit\'e du fluide seront
        parfaitement observ\'ees\,; car la partie post\'erieure \'etant
        (\textit{hyp.}) semblable et \'egale \`a la partie ant\'erieure\,, il est
        ais\'e de voir que les m\^emes valeurs de $p$ \& de $q$\,; qui donneront au premier
        instant l'\'equilibre \& l'incompressibilit\'e du fluide \`a
        la partie ant\'erieure\,, donneront les m\^emes r\'esultats \`a
        la partie post\'erieure. }}\\[-1.ex]

%

This statement is directly related to the remark in \S~70 of
d'Alembert's 1752 treatise. In fact, the assumption used by d'Alembert
in 1749 and 1752 to avoid a paradox is here lifted, since no separation
of streamlines occurs except at the needle-like end points.
D'Alembert here assumes that the solution with mirror symmetry is the
only one: ``The fluid has only one way to be
moved by the encounter of the body.'' The pressure forces at the front
and rear part of the body are then also axisymmetric and mirror
symmetric. Hence they
combine into a force of resistance (drag) which vanishes. D'Alembert concluded:

{\footnotesize
\noindent
	Thus I do not see, I admit, how one can satisfactorily explain
	by theory the resistance of fluids. On the contrary, it seems
	to me that the theory, developed in all possible rigor, gives,
	at least in several cases, a strictly vanishing resistance; a
	singular paradox which I leave to future Geometers to
	elucidate.\footnote{D'Alembert 1768:~138.  Je ne vois donc pas\,, je l'avoue\,,
	comment on peut expliquer par la th\'eorie, d'une maniere
	satisfaisante\,, la r\'esistance des fluides. ll me paro\^{\i}t au
	contraire que cette th\'eorie\,, trait\'ee \& approfondie avec toute
	la rigueur possible\;, donne\,, au moins en plusieurs cas\,, la
	r\'esistance absolument nulle\,; paradoxe singulier que je
	laisse \`a \'eclaircir aux G\'eometres [\textit{sic}]. }}\\[-1.ex]




It is clear that d'Alembert's argument is less general than that of Borda, since
he is restricting the formulation of the paradox to bodies with a head-tail 
symmetry. Nevertheless, d'Alembert is the first one to seriously
propose the vanishing of the drag as a paradox.

%

Twelve years later in  Volume~VIII of his ``Opuscules'' d'Alembert revisits
the paradox in the light of a letter received from ``a very great 
Geometer''
who is not named and who points out that, when considering the flow
inside or around a symmetric body, there may be, in addition to 
the symmetric solution, another one which does not possess such
symmetry and to which d'Alembert's argument for the vanishing of the resistance
does not apply.\footnote{D'Alembert, 1789:~212; Birkhoff, 1950:~21--22.} D'Alembert concurs
and discussed the issue at length. It should however be noted
that a breaking of the symmetry was already assumed by him
in his early work on the resistance when he assumed that 
the (hypothetical) points of attachment and detachment of the streamline
following the body are not symmetrically located 
(see Section~\ref{s:resistance}). 

Thus d'Alembert was definitely the first to formulate the vanishing
of the drag as a paradox within the accepted model of that  time,
namely  incompressible fluid flow, implicitly taken as ideal.\footnote{The idea
of  viscosity ripened only in the  XIXth century, see e.g. Darrigol,
2005; in the eighteenth century there was only a concept of
tenaciousness, e.g. resistance to the introduction of a body into
fluid, which was still a long way from actual viscosity.} He was
however formulating it only for bodies with head-tail symmetry, not
realizing that techniques introduced by Euler and Borda could have
allowed him to obtain the paradox for bodies of arbitrary shapes.

\section{Saint-Venant and the first precise formulation of the paradox (1846)}	
\label{s:venant}

In three notes published in 1846 and then in  a memoir published in
1847, Saint-Venant gives for the first time a general formulation of the paradox. A
detailed write-up, mostly dating from the same period, was published
only posthumously in 1888 and contains also a very interesting discussion
of previous work.\footnote{Saint-Venant, 1846, 1847, [1888].}
Saint-Venant's memoir marks the beginning of
the modern theory of the d'Alembert paradox which was to flourish, 
in particular with major contributions by Ludwig Prandtl.\footnote{Cf.,
  e.g. Darrigol, 2005:~Chap.~7.}.

We here give only a very brief description of the key results of
Saint-Venant.  He first
specified the properties of the incompressible fluid: the pressure force is normal to
the surface element on which it is acting and therefore equal in all
directions. The fluid moves steadily around a body at rest. He gives
a derivation of the paradox, closely related to Borda's.
Indeed, it 
suffices to establish the equation for the live forces acquired by the
fluid to see that the live force (energy) loss of the system
is zero:

{\footnotesize
\noindent
   If the motion has reached, as one always assumes, a steady state,
   the live force acquired by the system at every instant is zero; the
   work performed by the exterior pressures is also zero and the same
   applies to the work of the interior actions of the fluid whose
   density is assumed to be unchanging. Thus, the work of the impulse
   of the fluid on the body, and, consequently, the impulse itself,
   is necessarily equal to zero.\footnote{Saint-Venant,
   1847:~243--244.
Si le mouvement est arriv\'e,
   comme on le suppose toujours, \`a l'\'etat de permanence, la force
   vive, acquise \`a chaque instant par le syst\`eme, est nulle ; le
   travail des pressions ext\'erieures est nul aussi, et il en est de
   m\^eme du travail des actions int\'erieures du fluide dont nous
   supposons que la densit\'e ne change pas. Donc le travail de
   l'impulsion du fluide sur le corps, et, par cons\'equent, cette
   impulsion elle-m\^eme, est n\'ecessairement z\'ero.}}\\[-1.ex]

He adds that the situation is different for a real fluid made of
molecules
in which there is
friction at the contact between two neighboring fluid elements.
%

{\footnotesize
\noindent
	But one finds another result if, instead of an ideal fluid --
	object of the calculations of the geometers of the last
	century -- one uses a real fluid, composed of a finite
	number of molecules and exerting in its state of motion
	unequal pressure forces or forces having components tangential
	to the surface elements through which they act; components to
	which we refer as the friction of the fluid, a name which has
	been given to them since Descartes and Newton until
	Venturi.\footnote{Saint-Venant, 1847:~244. Mais on trouve un autre r\'esultat si, au
	lieu du fluide id\'eal, objet des calculs des g\'eom\`etres du
	si\`ecle dernier, on remet un fluide r\'eel, compos\'e de
	mol\'ecules en nombre fini, et exercant dans l'´\'etat du
	mouvement, des pressions in\'egales ou qui ont des composantes
	tangentielles aux faces \`a travers desquelles elles agissent~;
	composantes que nous d\'esignons par le nom de frottement du
	fluide, qui leur a \'et\'e donn\'e depuis Descartes et Newton
	jusqu'\`a Venturi.}}\\[-1.ex]

 Thus,
d'Alembert's paradox is explained by Saint-Venant for the
first time as a
consequence
of ignoring viscous forces. Of course, a precise  formulation of the paradox
would not have been possible without  a clear distinction between
ideal and viscous fluids.
%

\section{Conclusion}  
\label{s:conclusion}

The problem of the resistance of bodies moving in fluids was -- and still is
-- of great practical importance. It was thus naturally one of the first
non-trivial problems tackled within the nascent eighteenth century
hydrodynamics. Euler, who was not only a great ``Geometer'' but a person
acutely aware of the needs of gunnery and ship building, tried -- as we have
seen -- reaching beyond the old impact theory of Newton --- and failed. He
was lacking both the concept of viscous forces and a deep understanding of the
global aspects of the topology of the flow around a body. His ``failure'' --
as is frequently the case with major scientists -- was however very creative:
born was the idea of analyzing a steady flow into a set of fluid fillets of
infinitesimal and non-uniform section; he also managed to calculate the forces
acting on such fillet several years before there was any representation of the
dynamics in terms of partial differential equations. Borda, being both a
Geometer and an experimentalist, felt compelled to qualify as nonsensical a
very simple live-force argument discovered by himself and which predicted a
vanishing drag for bodies of arbitray shape. D'Alembert, another brillant
Geometer, was probably less constrained by experimental considerations, and
dared eventually to present the paradox known by his name. His proof reveals a
very good understanding of the global topology of the flow but otherwise is
very simple and limited intrinsically to bodies with a head-tail symmetry.

We must stress that the statement as a paradox is very much tied
to the type of analytical representation of an ideal flow. From this
point of view, experiments on flow past bodies,
be they real or thought experiments, have rather been an obstacle
to grasping the distinction  between an ideal fluid and a real one. The same
kind of epistemological obstacle has accompanied the earlier
birth of the principle of inertia, which no experiment could at that time
truly reveal; it was necessary to distance oneself from real
conditions 
and to find an appropriate mathematical representation. 
Finding such representations for fluid
dynamics was a painfully slow process: a full century elapsed
between Euler's fragmentary results on drag and Saint-Venant's
full understanding of the d'Alembert paradox.


\begin{acknowledgments}
Olivier Darrigol has been a constant source of inspiration to us
while we investigated the issues discussed here. We also thank
 Gleb K. Mikhailov for numerous remarks and  Rafaela Hillerbrand and
 Andrei Sobolevskii for 
their help.

\end{acknowledgments}


\end{document}